# Powered Multifocal THz Radiation by Mixing of Two Skew (ChG) Cosh-Gaussian Laser Beams in Collisional Plasma

Divya Singh[1,2] and Hitendra K. Malik[1*]

[1]*Plasma Waves and Particle Acceleration Laboratory, Department of Physics, Indian Institute of Technology Delhi, New Delhi – 110 016, India*

[2]*Department of Physics & Electronics, Rajdhani College, University of Delhi, New Delhi – 110 015, India*

*email-*dsingh@rajdhani.du.ac.in,

Considering a realistic situation, where electron-neutral collisions persist in plasma, analytical calculations are carried out for the THz radiation generation by beating of two Skew Cosh-Gaussian (SG) lasers of parameters n & s. The competency of these lasers over Gaussian lasers is discussed in detail with respect to the effects of collision and beam width on the THz field amplitude and efficiency of the mechanism. A critical transverse distance of the peak of the THz field is defined that shows a dependence on the skewness parameter of skew ChG lasers. Although electron-neutral collisions and larger beam width lead to the drastic reduction in the THz field when the skew ChG lasers are used in the plasma, the efficiency of the mechanism remains much larger than the case of Gaussian lasers. Moreover, the skewChG lasers produce stronger and multifocal THz radiation.

## I. INTRODUCTION

Terahertz radiation in the frequency range 0.1-10 THz lying between MW and IR regions have potential applications and utility in wide ranges of areas, because of which investigators are so keen to find out further possibilities[36] in this part of the spectrum. The THz radiations are being used for material characterization, medical imaging, tomography, topography, remote sensing[1], chemical and security identification[2, 3] etc. Now a day a variety of THz sources are commercially available, but these sources are large and relatively expensive to operate. Therefore, a lot of research is being carried out for achieving suitable THz sources. Faure *et. al.*[4] have proposed a scheme for the development of table top tunable THz source. Different groups around the globe have taken up different schemes for generating THz sources[5, 6]. Conventional schemes of the generation of THz sources involves the usage of electro-optic-crystal (EO crystal), such as ZnSe, GaP, $LiNbO_3$ or photo conductive antenna (PCA), super-luminous laser pulse interaction with large band gap semiconductors and dielectric[7−13]. Wang et al.[14] have observed an efficient THz radiation by mid IR few cycle pulses from gas targets. Ropagnol et al.[15] and Al-Naib et al.[16] have obtained intense THz radiation experimentally at low frequencies using an inter-digitated ZnSe large aperture PCA and also studied the effect of local field enhancement on the nonlinear THz response of a Si-based metamaterial. However, the limitation of using crystal is the material breakdown at high powers, low conversion efficiency and narrow bandwidth of emitted THz radiation. In order to overcome these limitations, plasma has been attempted as a nonlinear medium for the generation of THz radiation using highly intense ultra-short laser pulses in picosecond to femtosecond range. Our group has analytically investigated the THz radiation generation by tunnel ionization of a gas jet by employing superposed field of femtosecond laser pulses impinging onto it after passing through an axicon[17]. Cherenkov wake excited by a short laser pulse has been exploited theoretically by Yoshii et al.[18] to realize the THz radiation. Yugami et al.[19] have experimentally excited ion-wave wakefield by the resonant absorption of a short pulsed microwave with plasma and observed radiation from Cherenkov wakes in a magnetised plasma[20]. By focussing the fundamental and second harmonic lasers onto air, THz radiation has been obtained by Sheng et al.[21] via mode conversion of large amplitude plasma wake field excited in the presence of an axial density gradient. Manouchehrizadeh and Dorranian[22] have studied the effect of obliqueness of external magnetic field on the characteristics of magnetized plasma wakefield, which can be used for the THz radiation generation[23]. Antonsen et al.[24] have used a corrugated plasma channel for phase matched THz radiation generation by the ponderomotive force of a laser pulse. Malik et al.[25] have shown emission of Strong and collimated terahertz radiation by using super-gaussian lasers and spatial-triangular lasers in plasma. Patil[26] & Takale[27] had studied the ponderomotive effects on self-focussing in the interaction of cosh- Gaussian laser beams with a plasma. Lu et al.[28] studied the Propagation properties of cosh- Gaussian beams. Gill & Mahajan[29] also studied the Self-focussing of cosh-gaussian laser beam in a plasma. Many schemes have been discussed using Gaussian laser pulses impinged on plasma for THz generation. In our research paper we proposes the resonant excitation of THz radiation by mixing of two skew-ChG laser pulses with collisional plasma having significant value of electron neutral collision frequency. When it comes to application There are extensive applications of multifocal electromagnetic fields in the area of medicine and imaging. Due to non-invasive and non-destructive nature of THz fields, there may be several innovative and useful applications of multifocal THz radiations such as neuro-multifocal imaging, acousto-em spectroscopy, multifocal microscopy and various clinical applications. Persinger[30] studied the phenomenological consequences of exposure of complex em fields on human brain and observed neuroelectrical profiles. That helps treating neuro-epileptic disorders through multifocal neuroimaging, multifocal electro-retinography[31, 32]. US environment protection agency[33] in their review draft discussed Carcinogenicity treatment through multifocal Electromagnetic Fields. Kim et al[34] made the use of multifocal visual evoked potential (VEP) tests in the objective evaluation of the field in pediatric epilepsy surgery. Neeraj et al[35] studied lymphocytic vasculitis mimicking aggressive Multifocal Cerebral Neoplasm through MR Imaging and MR Spectroscopic appearances of multifocal fields. Bewersdorf et al[36] and Liu et al[37] used multifocal multiphoton microscopy (MMM) to present a real-time, direct-view multiphoton excitation fluorescence microscope that provides three-dimensional imaging at high resolution extensively used for bio-imaging. Xia et al[38] studied wide-field 2D multifocal optical-resolution photoacoustic-computed microscopy (2D-MFOR-PACM). Ruffini et al[39] Optimized multifocal transcranial current stimulation for weighted cortical pattern targeting from realistic modeling of electric Fields Recently to study brain disorders.

This is clear that many researchers have made use of plasma as a medium for THz radiation generation, but in most of the cases the electron-neutral collisions have not been taken into account. Since the collisions are very much integrated part of the plasma dynamics which cannot be ignored, we solve the problem of laser beating in a collisional plasma using SG laser beams for THz radiation generation. By comparing the results with the ones obtained by Gaussian lasers, we understand that the SG pulses although produce stronger radiation but are more susceptible to the collisions of electrons and neutrals in the plasma.

## II. CALCULATION OF THz FIELD

We consider a plasma where electron-neutral collisions (frequency ν) take place and which has space-periodically modulated density $N = N_0 + N_\alpha e^{i\alpha z}$ together with $N_\alpha$ as the amplitude and $\alpha$ the wave number of the density ripples. There are several techniques that may be used for adjusting the period and size of the density ripples [40–43]. The fields of the lasers are taken as $\vec{E}_j = E_{0L} \cosh^n\left(\frac{ys}{b_w}\right) \exp\left[-\left(\frac{y}{b_w}\right)^2\right] \exp[i(k_j z - \omega_j t)]\hat{y}$ together with $b_w$ as the beam width of the lasers and $j = 1,2$, where *s* is skewness parameter, n is the order of ChG beam such that *s=0 (Gaussian beam); s>0 (skew ChG beam)*

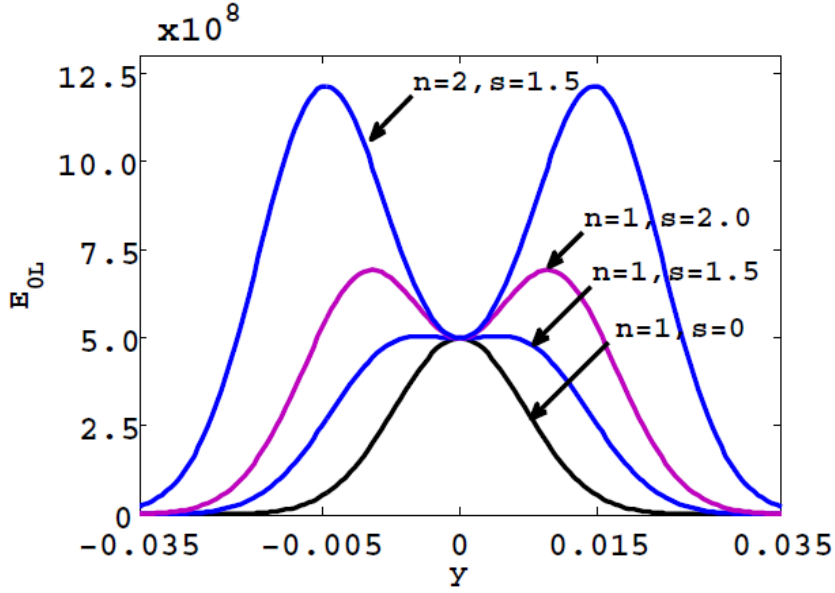

FIG.1. Beam profile of incident laser pulses is shown for skew-ChG beams for different orders-n and skewness parameter-s.

Since the plasma based schemes of THz radiation generation involving laser plasma interaction uses Gaussian beams. In the present paper we propose to make the use of skewChG beams as it is cleared their shape profile from the Fig-1. s=0 has a typical Gaussian shape whereas with s the shape of the beam changes. Remarkable dip at the peak intensity of the beam is observed for s=2.0 (n=1) and s=1.5 (n=2) i.e. critical parameters n & s of the beam.

In view of the field amplitude variation in the y-direction and laser beating, nonlinear ponderomotive force is realized at the frequency $\omega = \omega_1 - \omega_2$ and wave number $k = k_1 - k_2$. Initially the plasma electrons are under the action of laser fields, and the force acting on them (mass m) is expressed through equation of motion $m\frac{\partial \vec{v}_j}{\partial t} = -e\vec{E}_j - m\nu\vec{v}_j$ where collisional force is also taken into account. The laser fields impart oscillatory velocities to the electrons, given by $\vec{v}_j = \frac{e\vec{E}_j}{m(i\omega_j - \nu)}$. The nonlinear ponderomotive force of the lasers due to gradients in their fields is obtained as

$$\vec{F}_p^{NL} = -\frac{e^2 E_{0L}^2 \cosh^{2n}(ys/b_w)}{2m(i\omega_1 - \nu)(i\omega_2 + \nu)} \exp\left[-2\left(\frac{y}{b_w}\right)^2\right]\left[\left\{\frac{2ns}{b_w}\tanh(ys/b_w) - \frac{4y}{b_w^2}\right\}\hat{y} + ik\hat{z}\right]\exp[i(kz - \omega t)]$$

. (1)

Under the influence of the ponderomotive force, the electron oscillations become nonlinear, leading to the nonlinear density perturbations $N^{NL} = \frac{n_0}{mi\omega(\nu - i\omega)}\vec{\nabla}\cdot\vec{F}_p^{NL} = -\frac{\chi_e N_0 \vec{\nabla}\cdot\vec{F}_p^{NL}}{m\omega_p^2}$,

together with $\chi_e = -\frac{\omega_p^2}{\omega(\omega + i\nu)}$ and $\omega_p^2 = \frac{4\pi N_0 e^2}{m}$. Due to the nonlinear perturbations in electron density, some local field (space charge potential) is developed, which introduces the linear density perturbations $N^L = \frac{N_0 e \vec{\nabla}\cdot\vec{\nabla}\phi}{mi\omega(\nu - i\omega)} = -\frac{\chi_e \vec{\nabla}\cdot\vec{\nabla}\phi}{4\pi e}$. Now we calculate the resultant field or the force under the combined effect of linear and nonlinear density perturbations using Poisson's equation $\vec{\nabla}\cdot\vec{\nabla}\phi = 4\pi e(N^L + N^{NL})$ Hence

$$\vec{F}^L = e\vec{\nabla}\phi = \frac{\omega_p^2 \vec{F}_p^{NL}}{i\omega(1 + \chi_e)(\nu - i\omega)}.$$ (2)

The resultant nonlinear electron velocities under the action of linear force $\vec{F}^L$ and nonlinear force $\vec{F}_p^{NL}$ are obtained as $\vec{\upsilon}_y' = \frac{i\omega \vec{F}_{py}^{NL}}{m[i\omega(\nu - i\omega) - \omega_p^2]}, \vec{\upsilon}_z' = \frac{i\omega \vec{F}_{pz}^{NL}}{m[i\omega(\nu - i\omega) - \omega_p^2]}$, which yield the nonlinear oscillatory current density as

$$\vec{J}^{NL} = \vec{J}_y^{NL} = -\frac{1}{2}N'e\vec{\upsilon}^{NL} = -\frac{i\omega e N_\alpha e^{i\alpha z}\vec{F}_{py}^{NL}}{2m[i\omega(\nu - i\omega) - \omega_p^2]},$$ (3)

where $N' = N_\alpha e^{i\alpha z}$ and y subscript denotes y-component of current, $\vec{J}_y^{NL}$ would be suitable for the THz emission. This current oscillates at the frequency ω but its wave number k′ is different from the one of ponderomotive force. However, with the application of density ripples, the wave numbers can be tuned[44] and resonant excitation of the THz radiation can be realized through this nonlinear current. The field of the THz radiation (say $E_{THz}$) is governed by the following wave equation, which is obtained from the Maxwell's equations

$$-\nabla^2 \vec{E}_{THz} + \vec{\nabla}(\vec{\nabla}.\vec{E}_{THz}) = -\frac{4\pi i\omega}{c^2}\vec{J}^{NL} + \frac{\omega^2}{c^2}\varepsilon\vec{E}_{THz}.$$
(4)

Here $\varepsilon = 1 - \chi_e = 1 - \frac{\omega_p^2}{i\omega(\nu - i\omega)}$, and following the method used in Ref. 44, we obtain the normalized amplitude of the field of THz radiation as

$$\frac{E_{0THz}}{E_{0L}} = \frac{N_\alpha e E_{0L}\omega\omega_p^2 \exp^{-2\left(\frac{y}{b_w}\right)^2} \cosh^{2n}(ys/b_w)}{N_0 m b_w^2}\left\{y - \frac{nsb_w}{2}\tanh(ys/b_w)\right\} \times H$$

$$where\_H = \text{Re}\left[\frac{(\omega + i\nu)}{(\omega_1 + i\nu)(\omega_2 - i\nu)(\omega^2 - \omega_p^2 + i\omega\nu)^2}\right]$$
(5)

Here, it would be worth mentioning that this field is obtained only if the following phase matching condition is met

$$\text{Re}\left(\frac{\alpha c}{\omega_p}\right) = \frac{\omega}{\omega_p}\text{Re}\left[\left(1 - \frac{\omega_p^2}{\omega(\omega + i\nu)}\right)^{1/2} - 1\right].$$
(6)

this also means

$$\left(\frac{\alpha c}{\omega_p}\right) = \frac{\omega}{\omega_p} \times \left[\left(1 - \frac{\omega_p^2}{\omega^2 + \nu^2}\right)^{1/2} - 1\right].$$

Since quality of emitted THz radiations is determined by their amplitude, band width and efficiency. These are the key factors those are required to be achieved. Therefore analytical studies are made to study the nature of emitted THz radiations while Skew-ChG laser pulses of skewness parameter-s and order-n are copropagating in a collisional plasma of corrugated rippled density and further adequate optimization of parameters is needed to be done.

### III. EFFICIENCY OF THz RADIATION MECHANISM

The efficiency of THz radiation generation (say η) is defined as the ratio of energy of the THz radiation to the energy of the incident lasers, i.e. $\eta = \frac{\langle W_{THzE}\rangle}{\langle W_{LE}\rangle}$. Following Ref. [45], the energy density of the lasers i.e. the energy per unit volume, is calculated as $\langle W_{LE}\rangle = \frac{1}{8\pi}\varepsilon\frac{\partial}{\partial\omega}\left[\omega\left(1 - \frac{\omega_p^2}{\omega^2}\right)\right]\langle |E|^2\rangle$, while that of the THz field is $\langle W_{THzE}\rangle = \frac{1}{8\pi}\varepsilon\frac{\partial}{\partial\omega}\left[\omega\left(1 - \frac{\omega_p^2}{\omega^2}\right)\right]\langle |E_{THz}|^2\rangle$. Based on this, the total average energy densities are

evaluated and the efficiency $\eta$ of the THz radiation is obtained as

$$\eta = \left[ \frac{N_\alpha e E_{0L} \omega \omega_p^2 \exp^{-2\left(\frac{y}{b_w}\right)^2} \cosh^{2n}(ys/b_w)}{N_0 m b_w^2} \left\{ y - \frac{n s b_w}{2} \tanh(ys/b_w) \right\} \right]^2 \times$$

$$\left[ \frac{(\omega^2 + \nu^2)}{(\omega_1^2 + \nu^2)(\omega_2^2 + \nu^2)[(\omega^2 - \omega_p^2)^2 + \omega^2 \nu^2]^2} \right].$$

(7)

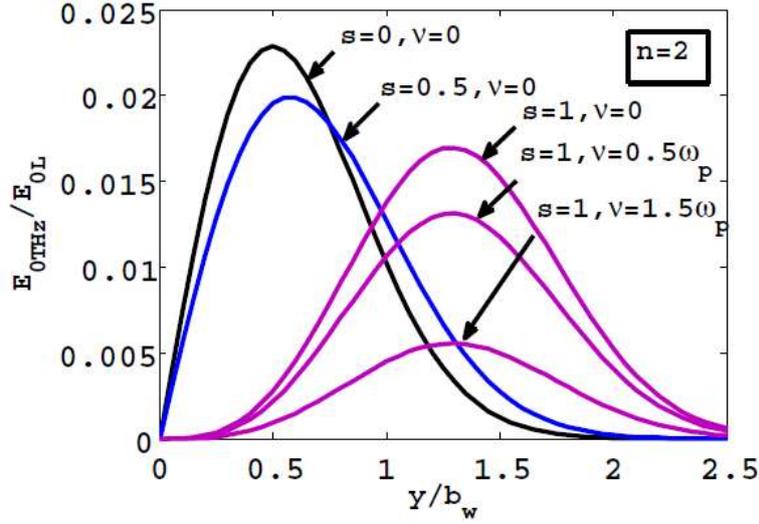

FIG. 2. Transverse profile of THz radiation field with different values of skewness parameter (below critical parameter) and collision frequency $\nu$, when $\omega_1 = 2.4 \times 10^{14}$ rad/s, $\omega_p = 2.0 \times 10^{13}$ rad/s, $E_0 = 5.0 \times 10^8$ V/m, $b_w = 0.01$ cm and $N_\alpha/N_0 = 0.1$.

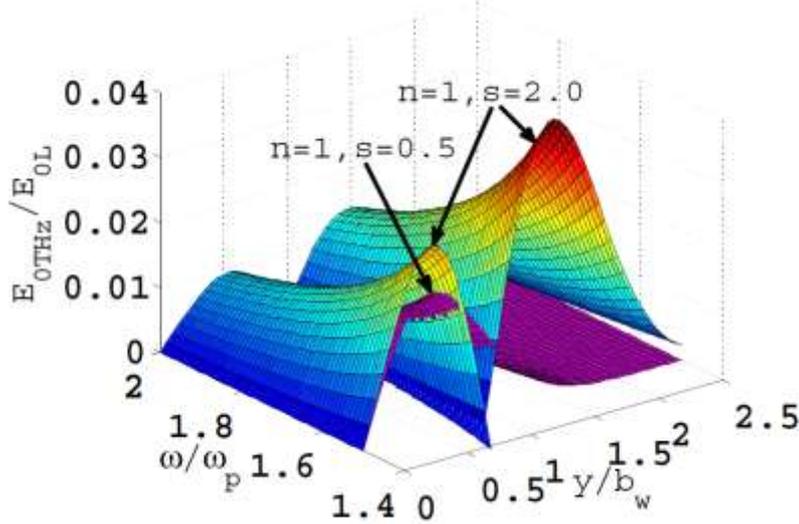

FIG.3. 3D surface plot of variation of normalized THz radiation field with normalized transverse distance y from the z- axis and normalised beating frequency (above critical parameter), for different values of s, $\omega_1 = 2.4 \times 10^{14}$ rad/s, $\omega_p = 2.0 \times 10^{13}$ rad/s, $E_0 = 5.0 \times 10^8$ V/m, and $b_w = 0.01$ cm.

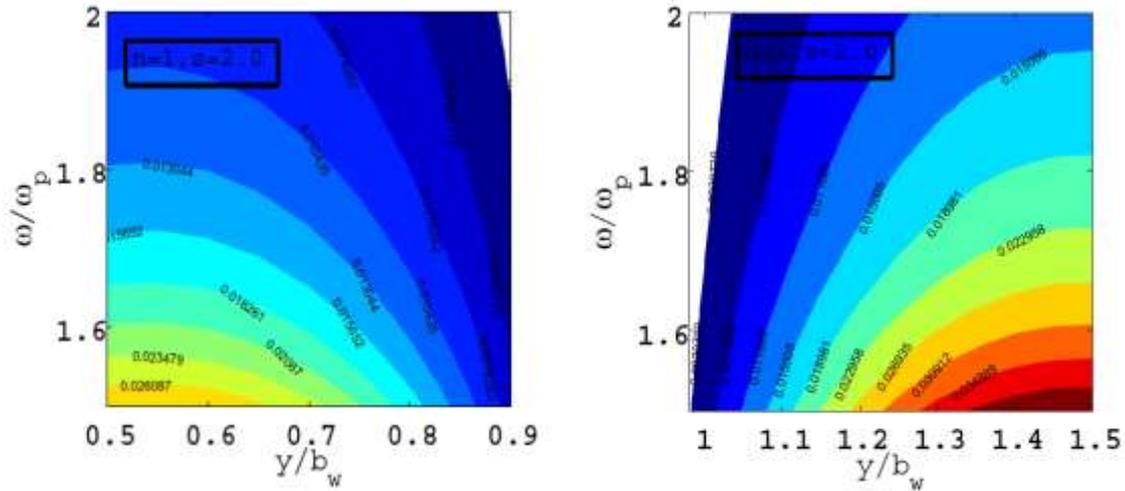

FIG. 4. 3D contour Plot (partwise $y/b_w$=0.5 to 0.9-Fig-a and 1.0 to 1.5-Fig-b)of variation in the normalized THz radiation field with normalized transverse distance y from the z- axis as well as normalised beating frequency (above critical parameters for n=1) for $\omega_1 = 2.4\times10^{14}$ rad/s, $\omega_p = 2.0\times10^{13}$ rad/s, $E_0 = 5.0\times10^8$ V/m, and $b_w = 0.01$ cm.

## IV. RESULTS AND DISCUSSION

From Fig-2 variation of normalised THz amplitude with normalised transverse distance is studied. For s = 0 i.e. the Gaussian laser beam gives THz amplitudes irrespective of the order of ChG beam same as obtained by Malik et al[34], but for increasing values of skewness parameter THz amplitude eventually falls but for a critical set of parameters (combination of n & s), the THz amplitude shows sudden enhancement as shown in 3D Fig-3. Moreover, the directionality of emitted THz is also get affected. The same effects are more elaborated in contour diagram Fig-4 depicting the THz amplitude to be found in both directions with enhanced amplitude. For low value of s=0.5, low THz amplitude is found in the direction of incident laser beam but for critical value of s=2.0, the enhanced high THz amplitude is found in complimentary direction of incident laser beam as well. This enhancement could be explained on the basis of the incident beam profile shape. Fig-1 explain the shape of the skew ChG laser pulse and for the combination of critical parameters of incident laser pulse a dip into the peak intensity for a specific value of transverse distance y=0 from z-axis is observed, splitting the incident laser pulse from top passing through and on interaction with gaseous plasma, produces electromagnetic radiations of high amplitudes. The frequencies of emitted amplitude may be controlled by the frequency of incident laser pulses. If incident laser frequencies lies in the range of pico or femtoseconds ($10^{12}/10^{14}$) the emitted radiations indeed will lie in terahertz range. Thus two closely placed intense THz peak amplitudes are reported for critical parameter of skew ChG laser beams and elsewhere very less intensities are found. As we suppose collisional plasma where electron neutral collisions are significant, leading to the decay of amplitude of the emitted THz due to thermal losses and departure of resonance condition wherein maximum transfer of energy could not take place, These results of collisional plasma are consistent with findings of Sharifian et al[37]. Since phase matching is the most important aspect of the entire process of THz emission which governs how, when and at what frequency the THz amplitude will emit. Due to the presence of electron neutral collisions, this condition shows variation in proportion with collisional frequency and accordingly the entire process get affected.

Figure 2 shows the transverse profile of electric field of THz radiation with a dependence on electron-neutral collision frequency ν. Clearly, in all the cases, the field attains a maximum value at a particular value of $y/b_w$ or y (for the fixed beam width $b_w$). This is attributed to the maximum magnitude of ponderomotive force at the said value of y for the fixed p and $b_w$ parameters. A comparison of all the graphs reveals that the emitted field profile is quite symmetrical and infers that the effect of collision frequency is much significant. Since the emitted radiation acquires lower field in the presence of larger collision frequency, It means the field of THz radiation falls at a faster rate with collision frequency.

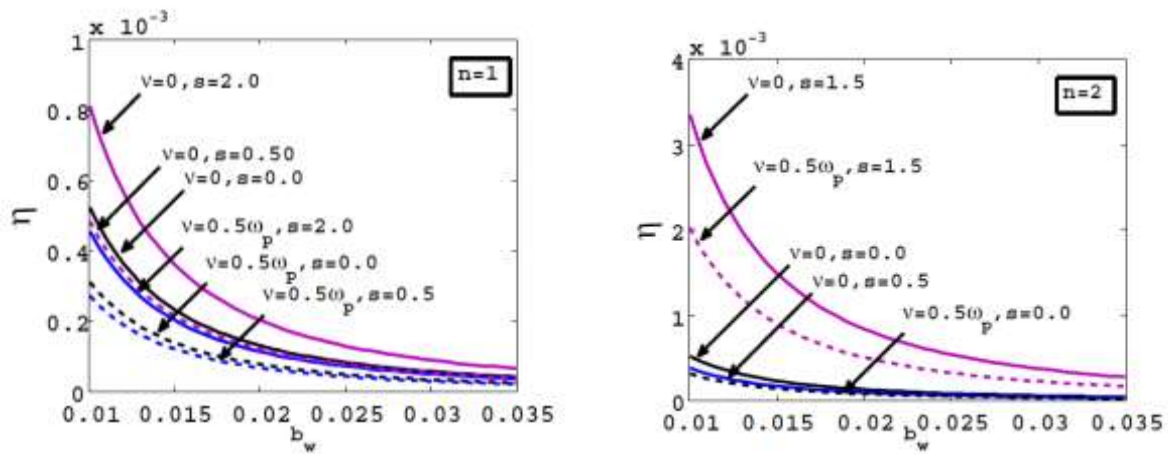

FIG. 5. Variation of efficiency of THz radiation mechanism with $b_w$ for different values of collision frequency and Skewness parameter for n=1 and n=2, when $\omega_1 = 2.4 \times 10^{14}$ rad/s, $\omega_p = 2.0 \times 10^{13}$ rad/s, $E_0 = 5.0 \times 10^8$ V/m, $y = 0.5 b_w$ and $N_\alpha/N_0 = 0.1$.

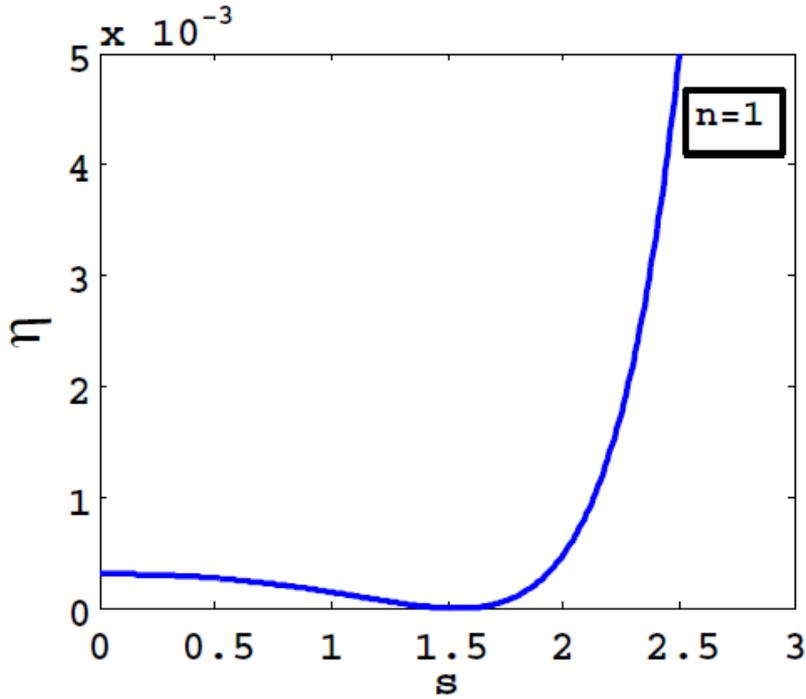

FIG. 6. Variation in the efficiency of THz radiation field with Skewness parameter

From Fig-5, efficiency is analysed for laser beam of variable beamwidth '$b_w$' since it is well known that the effect of beam width is very pronounced on the mechanism of generation of THz using plasma[44]. Lasers of smaller beam width helps in generation of higher amplitude THz therefore greatly enhances the efficiency of emitted THz. The effect of beam width is found to be remained unaltered even in the presence of electron neutral collisions prevailing plasma or if laser pulse of adequate skewness parameter are used. For very low collision frequency the effects on the efficiency of emitted THz are same as the non-collisional plasma whereas slight increase in the collision frequency upto the $1/10^{th}$ order of plasma frequency leads to a very sharp fall in the efficiency Fig-5. As we have seen how the amplitude of THz radiation is enhanced using skew ChG beams and hence the intensity also shows an enhancement for a particular combination of critical parameters. Collisions also have a significant decay effect on the efficiency but the same can be avoided using skew ChG beams of proper order (n) and skewness parameter (s). This effect can also be seen from Fig-6, that explains how efficiency shows significant change for changing skewness parameter – s.

A small change in beam width leads to the larger variation in the intensity gradient. We can also estimate the rate of decrease in efficiency with beam width in collisionless and collisional plasmas. The graphs marked with ν = 0 represent the case of collisionless plasma, as there is a very little change in efficiency (as per our calculation) till 1/100$^{th}$ of $\omega_p$ as the collision frequency. However, the graphs marked with ν = 0.5$\omega_p$ show a drastic reduction in the efficiency .

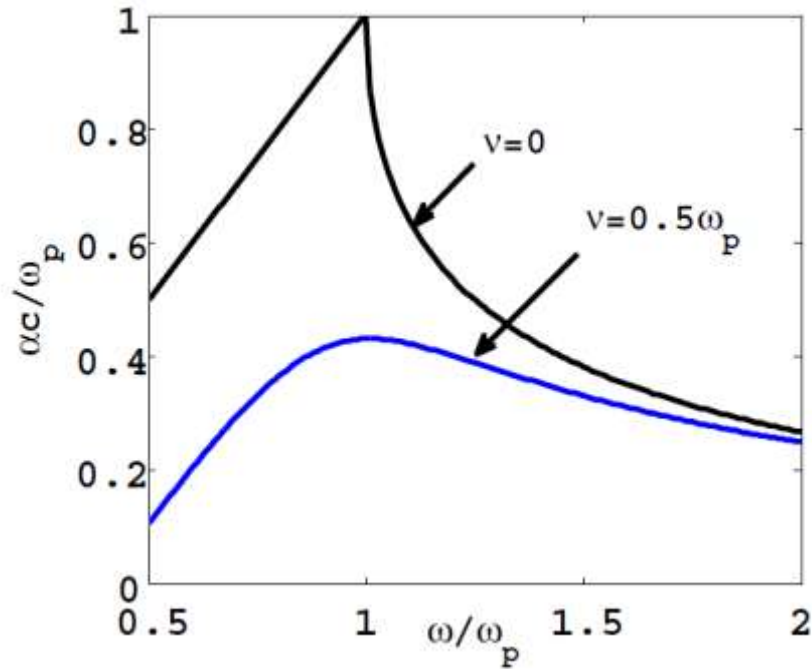

FIG. 7. Variation in the normalized wave number for periodic structure of plasma density ripples with normalized beating frequency

The present mechanism talks about the resonant excitation of THz radiation, when the condition (6) is satisfied. In order to match the wave numbers of ponderomotive force and nonlinear current, density ripples with periodicity 2π/α are required to be constructed in the plasma. Hence, αc/$\omega_p$ represents the normalized wave number corresponding to the density ripples. Figure 7 shows the variation of αc/$\omega_p$ with the normalized collision frequency ν/$\omega_p$ for different values of ω. This can be seen that α decreases with ν when ω > $\omega_p$, whereas it increases for the reasonable collision frequency when ω < $\omega_p$ . It means the distance between the density ripples should be adjusted accordingly. If ω > $\omega_p$ , then the separation should be larger between the density ripples, otherwise (for ω < $\omega_p$) these need to be constructed at smaller distances in order to get the efficient emission of THz radiation. If we look at the resonance condition, we find that the condition ω = $\omega_p$ obtained in the collisionless plasma is departed due to the electron-neutral collisions in the present case. Thus, the resonance condition reads $\omega = (\omega^2 - \nu^2)^{1/2}$. This also means that the frequency of the THz radiation in the collisional plasma is somewhat lower than the frequency obtained in the collisionless plasma. In view of this, the field of THz radiation shall attain lower amplitude in the collisional plasma. The decay in the field amplitude and hence, the efficiency is attributed to the energy and momentum loss involved in electron-neutral collision process. However, in the plasma, where $\nu \ll \omega$, this condition remains almost the same as in collisionless plasma.

In order to get adequate set of critical parameters n & s for maximum amplitude of THz radiation, we apply the principle of maxima and minima and we get the condition in the form of equation such as

$$1 + \frac{4nys}{b_w}\tanh\left(\frac{ys}{b_w}\right) - \frac{4y^2}{b_w^2} - n^2s^2\tanh^2\left(\frac{ys}{b_w}\right) - \frac{ns^2}{2}\mathrm{sech}^2\left(\frac{ys}{b_w}\right) = 0 \qquad (8)$$

Substitute $z = \tanh x$, where $x = \dfrac{ys}{b_w}$

Now the same equation can be written as

$$1 + 4nxz - \frac{4x^2}{s^2} - n^2s^2z^2 - \frac{ns^2}{2}\frac{\partial z}{\partial x} = 0$$

On solving this differential equation, we get proper sets of n & s giving appropriate value of transverse distance $y/b_w$ for maximum amplitude of emitted THz radiation.

$$\left(\frac{ns^2}{2} - n^2s^2\right)\tanh x = x - \frac{4x^3}{3s^2} - n^2s^2 x + 4n\int x\tanh x\, dx$$

For the approximation x<1 i.e. $\dfrac{ys}{b_w} < 1$. Above equation may be written in the form of cubic

$$\left(16n - \frac{8}{3s^2}\right)x^3 - 4nx^2 + (2 - ns^2)x + 4n = 0.$$

This is of the form $ax^3 + bx^2 + cx + d = 0$, Where coefficients are as follows

$a = 16n - \dfrac{8}{3s^2}, b = -4n, c = 2 - ns^2, d = 4n$.

The discriminant of the cubic equation is $\Delta = 18abcd - 4b^3d + b^2c^2 - 4ac^3 - 27a^2d^2$.

The general formula for the roots of cubic equation is

$$x_k = -\frac{1}{3a}\left(b + u_k C + \frac{\Delta_0}{u_k C}\right) \text{ where } u_k \text{ are the cubic roots of unity}$$

$$C = \sqrt[3]{\frac{\Delta_1 + \sqrt{\Delta_1^2 - 4\Delta_0^3}}{2}}, \Delta_0 = b^2 - 3ac, \Delta_1 = 2b^3 - 9abc + 27a^2d$$

$$u_1 = 1, u_2 = \frac{-1 + i\sqrt{3}}{2}, u_3 = \frac{-1 - i\sqrt{3}}{2}$$

Where $\Delta < 0$ (one real and two complex conjugate roots)

$\Delta = 0$ (multiple real roots exist)

$\Delta > 0$ (three distinct real roots exist)

Thus real root of cubic equation for the approximation we made is

$$\frac{y}{b_w} = -\frac{1}{3as}\left(b + \sqrt[3]{\frac{\Delta_1 + \sqrt{\Delta_1^2 - 4\Delta_0^3}}{2}} + \Delta_0 \left\{\sqrt[3]{\frac{\Delta_1 + \sqrt{\Delta_1^2 - 4\Delta_0^3}}{2}}\right\}^{-1}\right) \quad (9)$$

All the constants are already defined in terms of n & s. thus for getting an appropriate set of critical parameters, the value of y/b$_w$ can be tuned in order to get maximum THz radiation amplitude at desired position.

## V. CONCLUSIONS

Our analytical calculations show that the resonant excitation of THz radiation by mixing of two Skew ChG lasers in the presence of electron-neutral collisions is quite crucial, since these collisions involves thermal losses affect the entire process of THz generation results in the lesser efficiency than collisionless plasma, but the usage of skewChG laser beams helps improving higher amplitude and efficiency of emitted THz radiation. At resonance maximum energy transfer is found to take place by tuning of wave numbers of ponderomotive force and nonlinear current occurs with the help of wave number α of the density ripples. The density ripples at the closer distances help getting stronger THz radiation, whose field amplitude is further enhanced for the case of higher amplitude density ripples. Similar effects are observed for the smaller beam width and critical parameter n & s of the Skew ChG lasers.

## ACKNOWLEDGMENT

Divya Singh acknowledges Rajdhani College, New Delhi for giving NOC to conduct this research work at IIT-Delhi.

## REFERENCES

[1]B. Ferguson and X. C. Zhang, Nat. Mater. **1**, 26 (2002).

[2]Y. C. Shen, T.W. P. F. Today, B. E. Cole, W. R. Tribe, and M. C. Kemp, Appl. Phys. Lett. **86**, 241116 (2005).

[3]H. Zheng, A. Redo-Sanchez, and X. C. Zhang, Opt. Express **14**, 9130 (2006).

[4]J. Faure, J. V. Tilborg, R. A. Kaindl, and W. P. Leemans, Opt. Quan. Elec. **36**, 681 (2004).

[5]F. Blanchard, L. Razzari, H. Bandulet, G. Sharma, R. Morandotti, J. Kieffer, T. Ozaki, M. Reid, H. Tiedje, H. Haugen, Opt. Exp., **15**, 13212 (2007).

[6]P. Zhao, P. Ragam, S. Ding, Y. Zotova, Appl. Phys. Lett.,**101**, 021107, (2012)

[7]W. Shi, Y. J. Ding, N. Fernelius, and K. Vodopyanov, Opt. Lett. **27**, 1454 (2002).

[8]P. Zhao, S. Ragam, Y. J. Ding, and I. B. Zotova, Opt. Lett. **35**, 3979 (2010).

[9]Y. Jiang, D. Li, Y. J. Ding, and I. B. Zotova, Opt. Lett. **36**, 1608 (2011).

[10]D. Hashimshony, A. Zigler, and K. Papadopoulos, Phys. Rev. Lett. **86**, 2806 (2001); Appl. Phys. Lett. **74**, 1669 (1999).

[11]J. F. Holzman and A. Y. Elezzabi, Appl. Phys. Lett. **83**, 2967 (2003).

[12]G. H. Ma, S. H. Tang, G. K. Kitaeva, and I. I. Naumova, J. Opt. Soc. Am. B **23**, 81 (2006).

[13]E. Budiarto, J. Margolies, S. Jeong, J. Son, J. Bokor, IEEE J. Quan. Elec. **32**, 1839 (1996).

[14]W. M. Wang, S. Kawata, Z. M. Sheng, T. Y. Li and J. Zhang, Phys. Plasmas, **18**, 073108, (2011); Opt. Lett., **36**, 2608 (2011).

[15]X. Ropagnol, F. Blanchard, T. Ozaki and M. Reid, Appl. Phys. Lett., **103**, 161108 (2013); IEEE Photonics J., **3**, 174 (2011).

[16]I. Al-Naib, G. Sharma, M. Dignam, H. Hafez, A. Ibrahim, D. G. Cooke, T. Ozaki and R. Morandotti, Phys. Rev. B, **88**, 195203 (2013).

[17]H.K. Malik and A.K. Malik, Appl. Phys. Lett. **99**, 251101 (2011).

[18]J. Yoshii, C. H. Lai, C. Joshi, and W. B. Mori, Phys. Rev. Lett. **79**, 4194 (1997).

[19]N. Yugami, T. Higashiguchi, H. Gao, S. Sakai, K. Takahashi, H. Ito, Y. Nishida and T. Katsouleas, Phys. Rev. Lett. **89**, 065003 (2002).

[20]K. A. Hassan, M. Starodubtsev, H. Ito, N. Yugami and Y. Nishida, Phys. Rev. E, **68**, 364041 (2003).

[21]Z. M. Sheng, K. Mima, J. Zhang and H. Sanuki, Phys. Rev. Lett. **94**, 095003 (2005); Z. M. Sheng, K. Mima and J. Zhang, Phys. Plasmas **12**, 123103 (2005).

[22]M. Manouchehrizadeh and D. Dorranian, J. Theor. Appl. Phys. **7**, 43 (2013).

[23]S. Abedi, D. Dorranian, M. E. Abari and B. Shokri, Phys. Plasmas, **18**, 093108, (2011).

[24]T. M. Antonsen Jr., J. Palastra, and H. M. Milchberg, Phys. Plasmas **14**, 033107 (2007).

[25]H. K. Malik and A. K. Malik, Euro. Phys. Lett.,**100**, 45001(2012).

[26]S. D. Patil and M. V. Takale, Laser Phys. Lett. **10**, 115402 (2013).

[27]S. D. Patil, M. V. Takale, S. T. Navare, V. J. Fulari and M. B. Dongare, Optics & Laser Technology, **44**, 314-317 (2012).

[28]B. Lu, H. Ma and B. Zhang, Optics Comm.**164**, 165-170(1999).

[29]T. S. Gill, R. Mahajan and R. Kaur, Phys. of Plasmas **18**, 033110 (2011).

[30]M.A. Persinger, ISSN 0031-5125, Perceptual and Motor Skills, 80, 791-799, 1995.

[31]20 HKJ Ophthalmol Vol.15 No.1

[32]Hood DC, Zhang X., Ophthalmol. 2000;100(2-3):115-37.

[33]United States Environmental Protection Agency Office of Research and Development, Washington, DC 20460, EPA/600/6-90/005B External Review Draft October 1990

[34]Kim YJ, Yukawa E, Kawasaki K, Nakase H, Sakaki T. Neurosurg. 2006 Mar; 104 (3 Suppl):160-5.

[35]Neeraj J. Panchal, Soheil Niku, and Steven G. Imbesi, AJNR Am J Neuroradiol 26:642–645, March 2005

[36]Jorg Bewersdorf, Rainer Pick, and Stefan W. Hell, May 1, 1998 Vol. 23 No. 9 OPTICS LETTERS 655

[37]Lixin Liu, Lei Wang, Junle Qu, Ziyang Lin, Zhe Fu, Wenqing Liu, and Hanben Niu Euro. Conf. on Biomedical Optics, Munich, Germany, June 17, 2007 ISBN:9780819467720

[38]Jun Xia, Guo Li, Lidai Wang, Mohammadreza Nasiriavanaki, Konstantin Maslov, John A. Engelbach, Joel R. Garbow, and Lihong V. Wang Optics Letters, 38, 24, pp. 5236 (2013)

[39]Giulio Ruffini, Michael D. Fox, Oscar Ripolles,, Pedro Cavaleiro Miranda, Alvaro Pascual-Leone NeuroImage Volume 89, 1 April 2014, Pages 216–225

[40]K. Y. Kim, A. J. Taylor, T. H. Glownia, and G. Rodriguez, Nat. Photonics **153**, 1 (2008).

[41]C. C. Kuo, C. H. Pai, M. W. Lin, K. H. Lee, J. Y. Lin, J. Wang, and S. Y. Chen, Phys. Rev. Lett. **98**, 033901 (2007).

[42]S. Hazra, T. K. Chini, M. K. Sanyal, and J. Grenzer, Phys. Rev. B **70**, 121307(R) (2004).

[43]B. D. Layer, A. York, T. M. Antonson, S. Varma, Y.-H. Chen, Y. Leng, and H. M. Milchberg, Phys. Rev. Lett. **99**, 035001 (2007).

[44]A. K. Malik, H. K. Malik, and U. Stroth, Phys. Rev. E **85**, 016401 (2012).

[45]E. J. Rothwell and M. J. Cloud, *Electromagnetics*, **2nd ed.** (CRC/Taylor & Francis, London, 2009), p. 211.

[46]M. S. Sherwin, C. A. Schmuttenmaer, and P. H. Bucksbaum, Report of DOE-NSF-NIH Workshop on Opportunities in THz Science, 2004.

[47]M. Sharifian, H.R. Sharifinejad and H. Golbakhsi, J. Plasma. Phys., pp 1-11, Cambridge University Press (2014)